\documentclass[apl,reprint,superscriptaddress]{revtex4-2} 
\usepackage{gensymb}
\usepackage{notes2bib}
\usepackage{graphicx}
\usepackage{textcomp}
\usepackage{verbatim}
\usepackage{amsmath}
\usepackage{setspace}
\usepackage{multirow}

\usepackage[dvipsnames]{xcolor}

\usepackage[colorlinks=true,urlcolor=blue,linkcolor=blue,citecolor=blue,a4paper]{hyperref}
\graphicspath{{./images/}}

\begin{document}

\title{Magnetotransport properties in epitaxial films of metallic delafossite PdCoO$_2$: \\Effects of thickness and width variations in Hall bar devices}


\author{Arnaud P. \surname{Nono Tchiomo}\textcolor{blue}{\textsuperscript{\ddag}}}
\affiliation{Department of Physics and Astronomy, Louisiana State University, Baton Rouge, LA 70803, USA}
\author{Anand Sharma\textcolor{blue}{\textsuperscript{\ddag}}}
\affiliation{Department of Physics and Astronomy, Louisiana State University, Baton Rouge, LA 70803, USA}
\author{Sethulakshmi Sajeev}
\affiliation{Department of Physics and Astronomy, Louisiana State University, Baton Rouge, LA 70803, USA}
\author{Anna Scheid}
\affiliation{Max Planck Institute for Solid State Research, Heisenbergstr. 1, 70569 Stuttgart, Germany}
\author{Peter A. van Aken}
\affiliation{Max Planck Institute for Solid State Research, Heisenbergstr. 1, 70569 Stuttgart, Germany}
\author{Takayuki Harada}
\affiliation{Research Center for Materials Nanoarchitectonics (MANA), National Institute for Materials Science, Tsukubashi, Ibaraki 305-0044, Japan}
\author{Prosper Ngabonziza}
\email[corresponding author: ]{pngabonziza@lsu.edu}
\affiliation{Department of Physics and Astronomy, Louisiana State University, Baton Rouge, LA 70803, USA}
\affiliation{Department of Physics, University of Johannesburg, P.O. Box 524 Auckland Park 2006, Johannesburg, South Africa}

\date{\today}

\begin{abstract}
	
We report on a combined structural and magnetotransport study of Hall bar devices of various lateral dimensions patterned side-by-side on epitaxial PdCoO$_2$ thin films. We study the effects of both the thickness of the PdCoO$_2$ film and the width of the channel on the electronic transport and the magnetoresistance properties of the Hall bar devices. All the films with thicknesses down to 4.88~nm are epitaxially oriented, phase pure, and exhibit a metallic behavior. At room temperature, the Hall bar device with the channel width $\text{W}=2.5~\mu\text{m}$ exhibits a record resistivity value of 0.85~$\mu\Omega$cm, while the value of  2.70~$\mu\Omega$cm is obtained in a wider  device with channel width $\text{W}=10~\mu\text{m}$. For the 4.88~nm thick sample, we find that while the density of the conduction electrons is comparable in both channels, the electrons move about twice as fast in the narrower channel. At low temperatures, for Hall bar devices of channel width $2.5~\mu\text{m}$ fabricated on epitaxial films of thicknesses 4.88 and 5.21~nm, the electron mobilities of $\approx$~65 and 40~cm$^2$V$^{-1}$s$^{-1}$, respectively, are extracted. For thin-film Hall bar devices of width $10~\mu\text{m}$ fabricated on the same 4.88 and 5.21~nm thick samples, the mobility values of $\approx$~32 and 18~cm$^2$V$^{-1}$s$^{-1}$ are obtained. The magnetoresistance characteristics of these PdCoO$_2$ films are observed to be temperature dependent and exhibit a dependency with the orientation of the applied magnetic field. When the applied field is oriented 90$\degree$ away from the crystal \textit{c}-axis, a persistent negative MR at all temperatures is observed; whereas when the field is parallel to the \textit{c}-axis, the negative magnetoresistance is suppressed at temperatures above 150~K. 

\end{abstract}


\maketitle
The delafossite oxides, with general molecular formula of ABO$_2$ 
(A=Pd or Pt), have captured significant attention due to their extraordinary electronic  and structural properties. The key electronic  characteristics include, for example, PtCoO$_2$ having the highest conductivity per carrier of all known materials and PdCoO$_2$ exhibiting the longest electron mean free path ($l_e{_{(\text{T}=4 \text{ K})}}=  21.4 \,\mu $m) for all known  oxide  materials~\cite{CWHicks_2012,PKushwaha_2015,APMackenzie_2017}. The huge conductivity results from extremely broad conduction bands based on the 4$d$-5$s$ electrons of Pd, and the 5$d$-6$s$ electrons of Pt, whose character is nearly free electron like~\cite{APMackenzie_2017}. 

\begin{figure*}[!t]
	\centering 
	\includegraphics[width=0.75\textwidth]{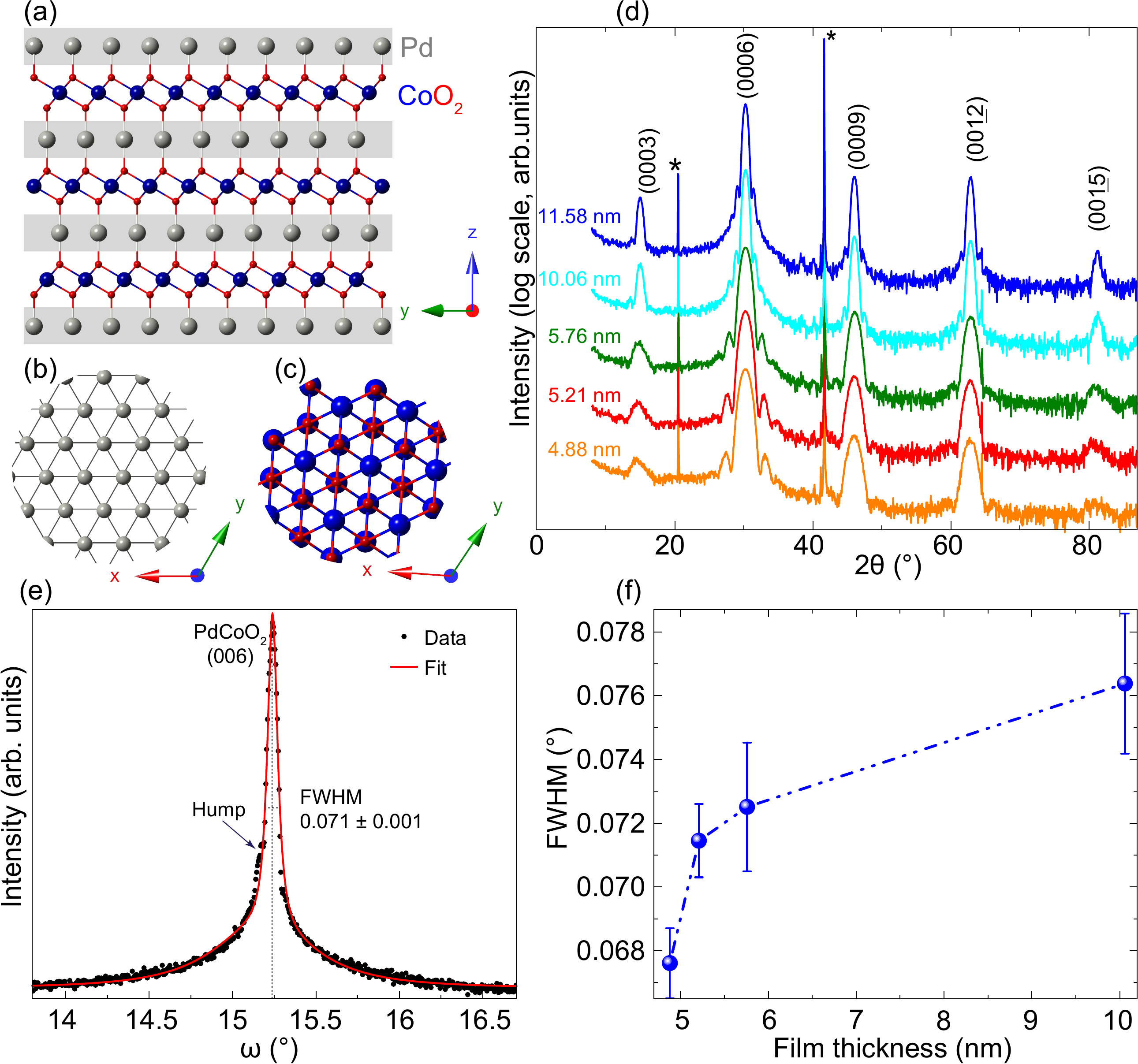}
	\caption{\textbf{Structural characterizations of unpatterned PdCoO$_2$ films}. (a) A (100) atomic plane view of the crystal structure of delafossite PdCoO$_2$ presenting alternating stacks of highly conducting Pd and insulating CoO$_2$ layers. The layers are connected through O--Pd--O dumbbells. Both layers form  triangular lattices as shown in the atomic (001) planes in (b) and (c), respectively for Pd and CoO$_2$. (d) Thickness-dependent $2\theta-\omega$ x-ray diffraction patterns of PdCoO$_2$ epitaxial films grown on $(0001)-$oriented Al$_2$O$_3$ substrates. The substrate peaks are indicated by asterisks (*) and the thickness of the films are indicated by the numbers adjacent to each XRD plot. (e) Rocking curve of a representative PdCoO$_2$ sample acquired around the main (006) film peak, and fitted to extract the full width at half maximum (FWHM). (f) Variation of the FWHM extracted from the (006) peak of the PdCoO$_2$ films shown in (d). Note the increase of the FWHM with increasing film thickness. \label{Fig1}}	 
\end{figure*}

We focus on the PdCoO$_2$ delafossite material.   The crystal structure of PdCoO$_2$ consists of two-dimensional (2D) Pd$^+$ and [CoO$_2$]-layers alternated along the \textit{c}-axis as shown in Fig.~\ref{Fig1}\textcolor{blue}{(a)}. The triangular coordinated Pd site layers are sandwiched between transition metal oxide layers in a stacking sequence [Fig.~\ref{Fig1}\textcolor{blue}{(b)}~-~\ref{Fig1}\textcolor{blue}{(c)}]. The 2D Pd$^+$ layers are electrically conductive, whereas the [CoO$_2$]- layers are electrically insulating. This quasi-2D layered crystal structure results in considerable anisotropy in electrical conduction. In single crystal samples, the room temperature \textit{ab}-plane resistivity of $\rho_{\text{ab}}\approx0.0026\text{ m}\Omega~\text{cm}$ and \textit{c}-axis resistivity of $\rho_{\text{c}} \approx 1.07 \text{m}\Omega~\text{cm}$ were reported~\cite{APMackenzie_2017}.

For fundamental research perspective, the combination  of above  interesting electronic characteristics together with its fascinating structural  properties  makes PdCoO$_2$ a promising material candidate for the investigation of fascinating and rich physics of the delafossite materials. Additionally, for technological applications, the high stability and mechanical rigidity of PdCoO$_2$ make it a promising electrode material for wide-bandgap semiconductor devices~\cite{THarada_2019}. 

Although PdCoO$_2$ and related   metallic   delafossites were first synthesized in 1971~\cite{R_Shannon_1971,CTPrewitt_1971}, PdCoO$_2$ was largely ignored for decades.  It is only in the mid 1990s that Tanaka and co-workers reported the growth of PdCoO$_2$ and PtCoO$_2$ crystals~\cite{MTanaka_1997}, as well as the first measurement of the temperature-dependent resistivity of PdCoO$_2$~\cite{,MTanaka_1996}. This brought a renewed interest in PdCoO$_2$ material for studying its basic properties like the thermoelectric power~\cite{MHasegawa_2002,OP_Khuong_2010,MEGruner_2015}, electronic structure~\cite{KKim_2009,VEyert_2008,KPOng_2010} and high anisotropy in structure, conductivity and compression behavior~\cite{RDaou_2015,HJNoh_2009,MHasegawa_2003}. However, the   physics   of   PdCoO$_2$ has  been  primarily  studied by using  single  crystal samples. Despite  decades  of  research, these single crystals are still  limited  in  size of $\approx$~3~mm  in diameter~\cite{RDShannon_1971,THiroshi_2007,PKushwaha_2015}. Thus, to allow further studies of its physical properties, particularly  as  its  thickness  is  decreased  down  to  few unit cells,  and  the  assessment  of  proof-of-principle  spintronic  devices, thin film samples with  large  area  and  smooth  surfaces are desired. 

Recently, the thin films of metallic delafossites with thicknesses down to a few nanometers have been reported by several groups~\cite{THarada_2020,THarada_2018,MBrahlek_2019,PYordanov_2019,SunJiaxin_2019,THarada_2019,WRenhuai_2020}. These thin films were grown along the $c$-axis direction on substrates with pseudo-triangle lattices, such as Al$_2$O$_3$(0001)~\cite{KTomoki_2023,QiSong_2022,QiSong_2024} and $\beta-$Ga$_2$O$_3$ $(\bar{2}01)$~\cite{THarada_2019,THarada_2020-01,THarada_2020-02}. The Al$_2$O$_3$ (0001) substrates have been mostly used for the thin-film growth of metallic delafossites~\cite{THarada_2021-01}. Specifically, epitaxial PdCoO$_2$ films have been synthesized  using  sputtering~\cite{PFCarcia_1980,THarada_2023}, pulsed-laser  deposition  (PLD)~\cite{THarada_2018,PYordanov_2019,THarada_2021,THarada_2018}, molecular-beam  epitaxy  (MBE)~\cite{MBrahlek_2019,QiSong_2024,SunJiaxin_2019,QiSong_2022} and solid-phase reactions of precursors~\cite{WRenhuai_2020}.

The magnetoresistance (MR) properties of PdCoO$_2$ have been characterized both in bulk single crystals~\cite{APMackenzie_2017} and in epitaxially-grown thin film prepared mostly on sapphire substrates~\cite{WRenhuai_2020,THarada_2018,PYordanov_2019,THarada_2021,THarada_2018,PFCarcia_1980,THarada_2023,MBrahlek_2019,QiSong_2024,SunJiaxin_2019,QiSong_2022}. The MR data were only  reported  for a single (one) Hall bar device that was structured by focused ion beam on PdCoO$_2$ single crystals~\cite{NNabhanila_2018}, and also for a single Hall bar device fabricated on epitaxial PdCoO$_2$ films~\cite{lee2021nonreciprocal,THarada_2021,QiSong_2022}. However, patterning several Hall bar devices of various channel width side-by-side on the same PdCoO$_2$ sample offers an advantage over single crystal-based devices as it provides the opportunity to perform comparative study of magnetotransport properties from the same sample, measured in similar conditions~\cite{ngabonziza2024magnetotransport}. Moreover, this practice is ideal for the exploration of lateral dimensional confinement effects in epitaxial films for tuning their electronic ground states.

In this paper, we report on a combined structural and magnetotransport study of Hall bar devices of various lateral sizes patterned side-by-side on the same PdCoO$_2$ thin film. In magnetotransport, we focus on exploring the effects of the film's thickness (t), the variation of the width (W) in Hall bar devices as well as the orientation of the magnetic field \textit{\textit{B}} on the MR characteristics. We use pulsed laser deposition (PLD) for the epitaxial growth of these thin films. The growth conditions are optimized to achieve phase-pure metallic delafossite PdCoO$_2$ thin films of different t  down to a few manometers. Subsequent structural analyses using x-ray diffraction (XRD) for unpatterned epitaxial films and high-resolution scanning transmission electron microscopy (STEM) for patterned Hall bar devices, confirm that these films and thin-film devices are epitaxially oriented and phase pure. The temperature dependence  of  the  in-plane  resistivity  as  a  function  of  film  thickness show metallic behavior down  to $\sim$~4.88~nm. 

For metallic delafossite PdCoO$_2$ of various t, we fabricated side-by-side on the same PdCoO$_2$ films several Hall bars of various W, and then explored their magnetotransport properties. In particular, for two Hall bar devices of channel width $\text{W}=2.5$ and $10~\mu\text{m}$, we found that the electron mobilities ($\mu_e$), simultaneously extracted from Hall measurements at various temperatures, changed systematically as W and t were varied. At 2~K, we extracted $\mu_e$ values of $\approx$~65 and 40~cm$^2$V$^{-1}$s$^{-1}$ from the Hall bar devices of $\text{W}=2.5~\mu\text{m}$ fabricated on epitaxial films of $\text{t}=4.88$  and 5.21~nm, respectively; whereas the $\mu_e$ values of $\approx$~32 and 18~cm$^2$V$^{-1}$s$^{-1}$ were obtained from the devices with channel $\text{W}=10~\mu\text{m}$ fabricated on the same respective films. In addition, we find that the magnetotransport is strongly temperature dependent, and present a negative MR that can persist at all temperatures or switch to positive MR at certain temperatures depending on the orientation of the applied magnetic field with respect to the sample \textit{c}-axis.

\begin{figure*}[!t]
	\centering 
	\includegraphics[width=0.7\textwidth]{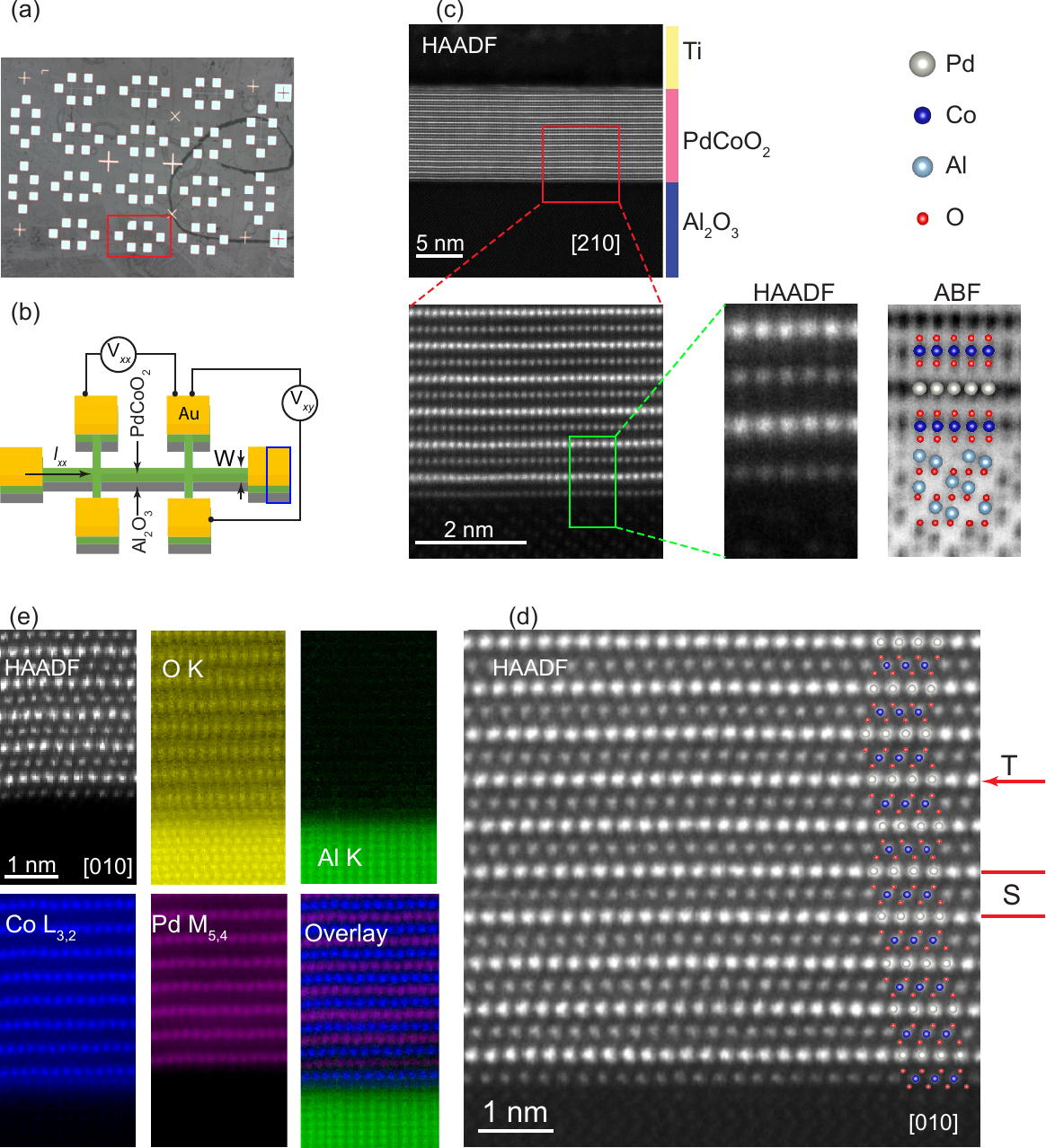}
	\caption{\textbf{Device fabrication and Microstructural characterization of the PdCoO$_2$ sample grown on (0001)-oriented Al$_2$O$_3$ substrate using PLD}. (a) Optical micrograph image showing a top view of a whole PdCoO$_2$ sample on which several Hall bar devices of different channel widths (W) were fabricated side-by-side. The channel widths of these devices are W~= 10~$\mu$m, 5~$\mu$m, 2.5~$\mu$m, 1~$\mu$m and 500~nm. (b) Schematic illustration of a typical Hall bar device (red rectangle in (a)) indicating the transport measurement configuration. (c)  Cross-sectional HAADF-STEM image obtained from a Hall bar device as indicated by the blue rectangle in (b). The enlarged insets show the atomic-scale structure of the delafossite-substrate interface along the [210] orientation of the film with an structural overlay of the atomic models on simultaneously acquired HAADF and ABF images. The HAADF STEM images provide strong contrast for the heavier elements Pd and Co, while the O and Al atomic columns are only visible in the ABF images capturing electrons scattered toward lower angles. (d) HAADF STEM image along the [010] orientation of the film reveals domains with opposing orientations of the CoO$_6$ octahedra in subsequent $[\text{CoO}_2]^{-}$ layers indicating the presence of twins T (red arrow), and stacking faults S. (e) EELS elemental mapping across the PdCoO$_2$/Al$_2$O$_3$ interface obtained by extracting the respective edge signals from a 2D spectrum image across the [010] zone axis, indicating a sharp film-substrate interface. \label{Fig2}}	 
\end{figure*}

Epitaxial PdCoO$_2$ films were deposited on the (0001)-oriented  Al$_2$O$_3$ ($c-$Al$_2$O$_3$) at a substrate temperature of T$_\text{sub}=700~\degree$C under an oxygen pressure of $100-150$~mTorr in a PLD chamber using the 4$^{\text{th}}$ harmonic (266~nm) of Nd:YAG laser for ablation. PdCoO$_2$ and mixed-phase PdO$_{\text{x}}$ targets were alternately ablated to obtain a stoichiometric composition. We prepared films of different thicknesses varying from 4.88 to 11.58~nm. The thickness of the PdCoO$_2$ films was determined using the thickness fringes around the PdCoO$_2$ (0006) peak in the XRD and the x-ray reflectivity data~\cite{THarada_2020}. The electronic transport properties were measured by using a Quantum Design Physical Property Measurement System (PPMS), where a excitation current of $I=1~\mu$A was applied to the Hall bars. The MR properties of the PdCoO$_2$ films were studied by changing the orientation of the magnetic field \textit{B} such that B~$\parallel$~\textit{c}, and \textit{B} oriented 45$\degree$ and 90$\degree$ away from \textit{c}. Details on the Hall bar device fabrication are provided in the supplemental material. 

Figure~\ref{Fig1}\textcolor{blue}{(d)} depicts XRD $2\theta-\omega$ scans of PdCoO$_2$ films of different thicknesses. This thickness-dependent analysis exhibits the crystal structure purity of the grown samples. The film reflections form only parallel $(000l)$ planes of the bulk crystal structure of PdCoO$_2$, indicating an epitaxial and single phase growth along the \textit{c}-axis. Additionally, the clear Laue oscillations observed around the film peaks suggest that the films have smooth surfaces. Note that as the thickness of the films increases (increasing unit cells), the films diffraction peaks become sharper. All these observations are consistent with previous reports on the epitaxial growth of PdCoO$_2$ thin films using  PLD~\cite{THarada_2018}, MBE~\cite{SunJiaxin_2019,QiSong_2022,QiSong_2024,MBrahlek_2019}, and sputtering~\cite{THarada_2023}.

\begin{figure*}[!t]
	\centering 
	\includegraphics[width=0.7\textwidth]{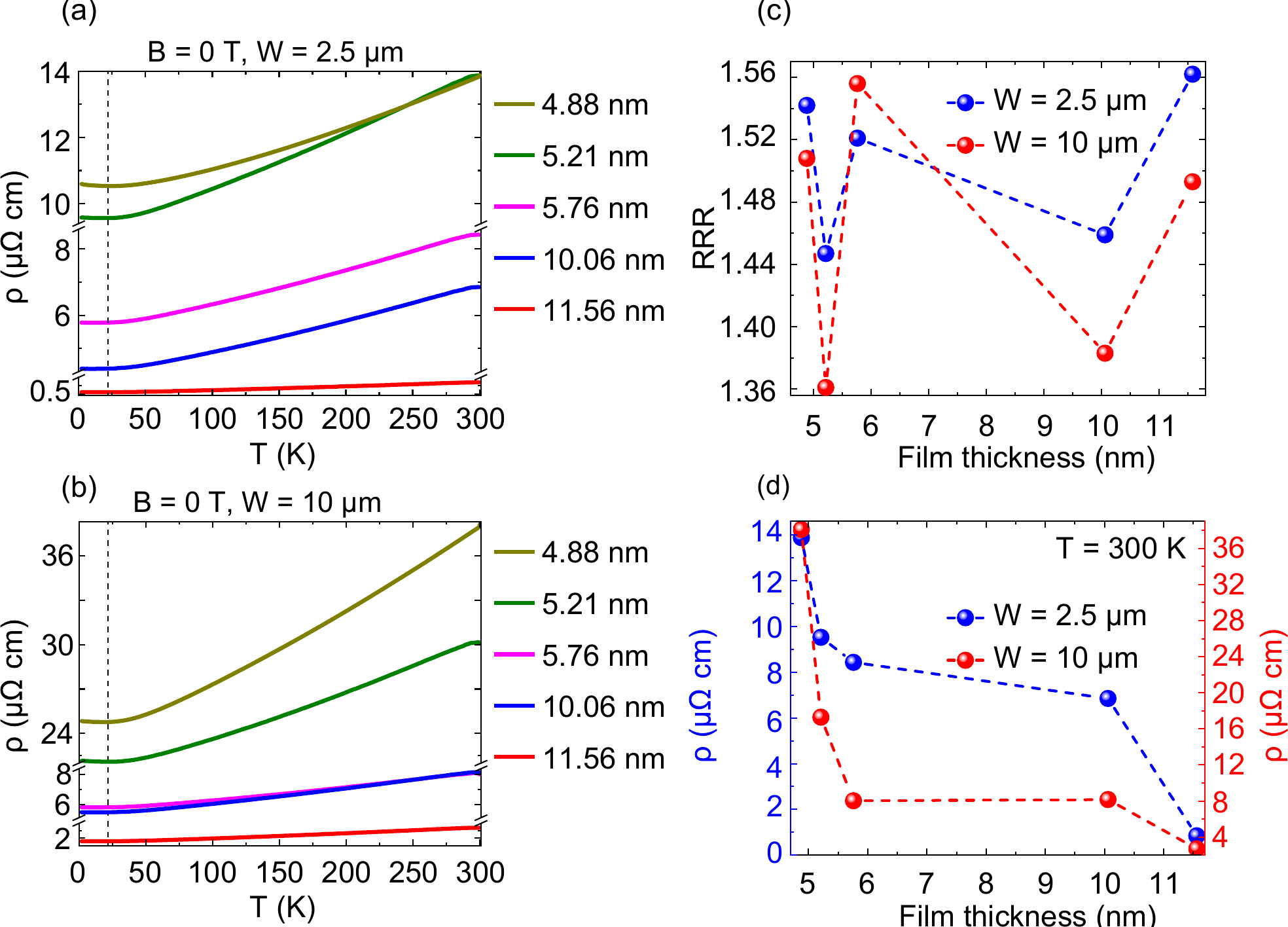}
	\caption{\textbf{Thickness-dependent electronic transport characteristics of PdCoO$_2$ epitaxial films}. (a) and (b) Temperature dependent resistivity at zero magnetic field for patterned PdCoO$_2$ Hall bar thin-film devices  for a range of thicknesses. The channel widths are (a) $\text{w}=2.5\,\mu$m and (b)  $\text{w}=10\,\mu$m. The vertical dashed lines indicate $T_{\text{min}}=20$~K. The patterned Hall bar devices in (a) and (b) were fabricated side-by-side on the same PdCoO$_2$ films in which electronic transport properties were studied. (c) Plot of the variation of the residual resistivity ratio (RRR) as $\rho_{\text{300~K}}/\rho_{\text{2~K}}$ VS films thickness for both Hall bar widths. (d) Variation of the room temperature (RT) resistivity values as a function of the film's thickness for both Hall bar geometries. The RRR and the RT resistivity values plotted in (c) and (d), respectively, are extracted from (a) and (b). Note that the overall trend with respect to both Hall bar device widths is that the confinement of charge carriers in a narrower channel results in improved electronic properties, as the scattering pathway is reduced. \label{Fig3}}	 
\end{figure*}

We studied the structural perfection of the samples by performing symmetric rocking curve $\omega$ scans  for the (0006) film peak as shown in Fig.~\ref{Fig1}\textcolor{blue}{(e)}. The full width at half maximum (FWHM) values extracted by fitting the rocking curves are plotted in Fig.~\ref{Fig1}\textcolor{blue}{(f)}. These small FWHM values hint at highly oriented PdCoO$_2$ films exhibiting minor out-of-plane misorientation~\cite{THarada_2018,SunJiaxin_2019,acharya2021twinned}.  It is noteworthy that other works on the heteroepitaxial growth of PdCoO$_2$ have reported azimuthal mosaicity (in-plane orientation distribution) by recording $\phi$ scans on asymmetric reflections. It was found that the PdCoO$_2$ films exhibit far greater mosaicity in the plane, and grow by forming twin domains which are rotated 180$\degree$ from one another [see Fig.~\textcolor{blue}{S1} of the supplemental material] and 30$\degree$ from the Al$_2$O$_3$ substrate~\cite{THarada_2018,SunJiaxin_2019,THarada_2023,MBrahlek_2019}. Nevertheless, we consistently noted a small hump in the line shape of the main peak in the rocking curves of all the films [see Fig.~\textcolor{blue}{S2} of the supplemental material]. This could be an indication of an adjacent crystalline domain having a different orientation to the main crystalline block~\cite{dolabella2022lattice}. Our Scanning transmission electron microscopy (STEM) measurements provide evidence of twin boundaries in the PdCoO$_2$ film as discussed in the following.

\begin{figure*}[!t]
	\centering 
	\includegraphics[width=0.85\textwidth]{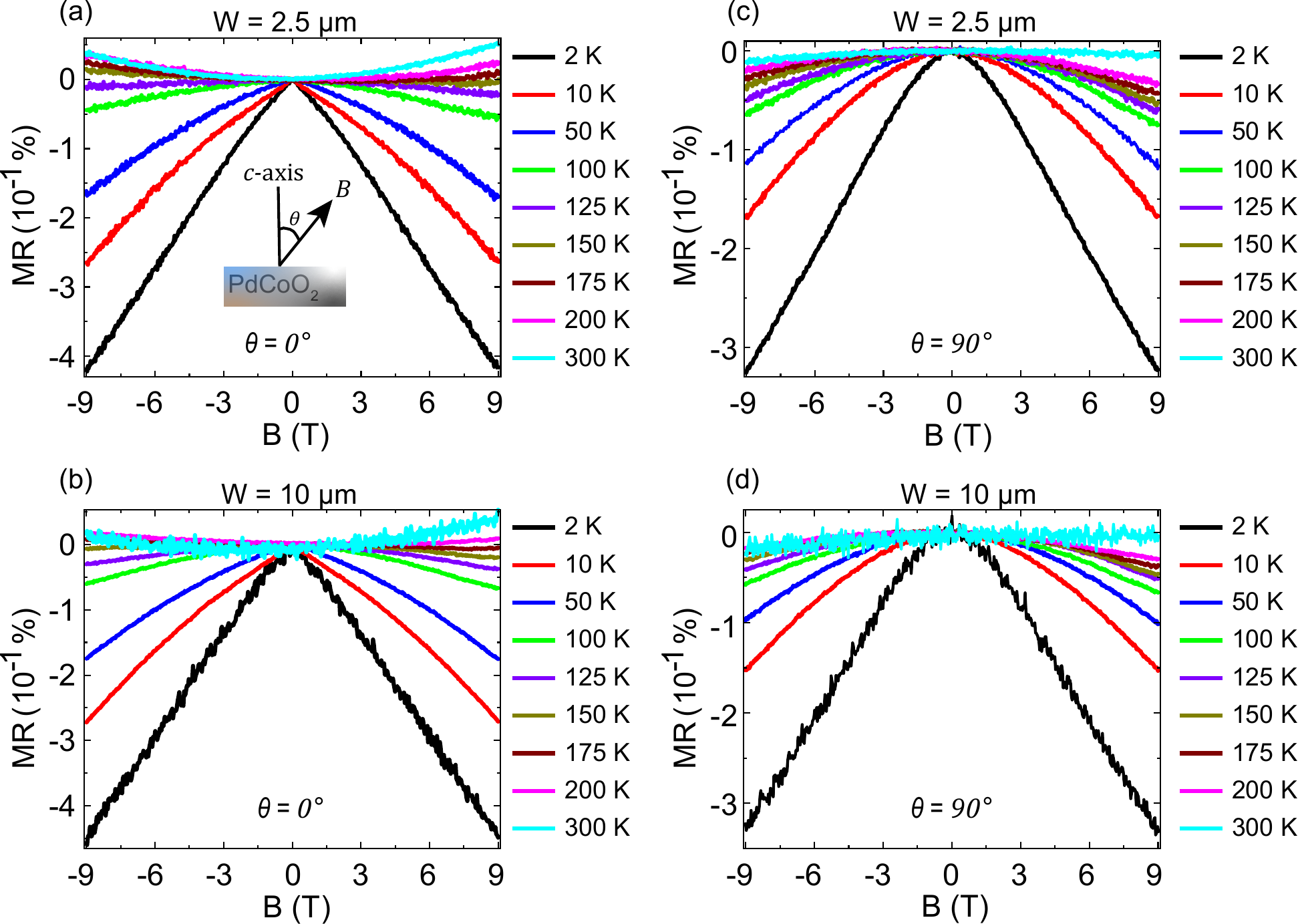}
	\caption{\textbf{Temperature-dependent magnetoresistance (MR) properties of epitaxial PdCoO$_2$ films measured at different \textit{B} orientations}. The data are from two Hall bar devices of channel widths $\text{W}=2.5~\mu$m and $\text{W}=10~\mu$m fabricated on the PdCoO$_2$ sample with thickness $\text{t}=4.88~\text{nm}$. (a) and (b) MR simultaneously acquired when $B\parallel c$, ie, $\theta=0\degree$. (c) and (d) MR simultaneously acquired when $B\perp c$, ie, $\theta=90\degree$.  The insert in (a) shows the orientation of \textit{B} with respect to the \textit{c}-axis. Note that when $\theta=0\degree$, the MR present a cross-over from negative to positive at temperatures above 150~K; but when $\theta=90\degree$, the negative MR persist at all temperatures. Note that through out the magnetotransport measurements, \textit{B} is always perpendicular to \textit{I}. \label{Fig4}}	 
\end{figure*}


Figure~\ref{Fig2}\textcolor{blue}{(c)} presents the high-angle annular dark field (HAADF) STEM images of the PdCoO$_2$ film along the [210] orientation, obtained from a cross-section of a Hall bar device, as illustrated in Fig.~\ref{Fig2}\textcolor{blue}{(b)}. The magnified HAADF image reveals a sharp and smooth interface between the PdCoO$_2$ film and the  Al$_2$O$_3$ substrate, consistent with the well-defined Laue oscillations observed in the XRD patterns [see, Fig.~\ref{Fig1}\textcolor{blue}{(d)}]~\cite{SunJiaxin_2019,scheid2025unveiling}. Furthermore, the periodic layer stacking of the PdCoO$_2$ film confirms the presence of a pure delafossite phase. The crystallographic structures of both the film and substrate can be overlaid on the simultaneously acquired HAADF and annular bright-field (ABF) images, which provide contrast for the lighter elements Al and O. 

The STEM images indicate that the epitaxial relationship between the film and substrate is established through the initial $[\text{CoO}_2]^{-}$ layer  [Fig.~\ref{Fig2}\textcolor{blue}{(c)}], which provides the low-energy interface for contact with the substrate, resulting in a stable heterointerface~\cite{SunJiaxin_2019,ok2020pulsed,scheid2025unveiling}. This is commonly achieved in the growth of delafossite  PdCoO$_2$,  PtCoO$_2$ and  PdCrO$_2$ materials on the sapphire substrate using both PLD and MBE~\cite{SunJiaxin_2019,QiSong_2022,QiSong_2024,THarada_2018}. Moreover, the HAADF STEM images along the [010] orientation [Fig.~\ref{Fig2}\textcolor{blue}{(d)}] reveal domain boundaries, indicating the presence of stacking faults (S) and twin boundaries (T) in the film. The stacking faults correspond to the translational displacement of oxygen octahedra in successive delafossite layers; and, the twin boundaries depict the mirrored orientation of the oxygen octahedra in adjacent layers. These features are commonly observed in delafossite thin films grown on \textit{c}-axis-oriented substrates~\cite{SunJiaxin_2019,THarada_2018,MBrahlek_2019,roudebush2013structure}.

EELS elemental mapping across the PdCoO$_2$/Al$_2$O$_3$ interface, shown in Fig.~\ref{Fig2}\textcolor{blue}{(e)}, reveals the atomic elemental distribution along the [010] orientation of the film. The extraction of element-specific energy loss edge signals enables the mapping of the delafossite structure composition. The absence of intermixed layers at the interface suggests that the film exhibits good stoichiometry~\cite{nono2022combined}, and confirms its epitaxial growth. The EELS elemental maps further verify that the film nucleates with a $[\text{CoO}_2]^{-}$ layer.


The PdCoO$_2$ samples also exhibit good electrical characteristics over the varying thicknesses as portrayed in the temperature dependence of the resistivity at zero magnetic field presented in Fig.~\ref{Fig3}, for the Hall bar devices with channel widths $2.5$ and $10~\mu\text{m}$. The resistivity is seen to decrease with increasing film's thickness. For each sample, the resistivity displays a positive temperature coefficient, consistently for the two Hall bar devices [Fig.~\ref{Fig3}\textcolor{blue}{(a)} and \ref{Fig3}\textcolor{blue}{(b)}]. This is indicative of a metallic behavior in all the films over almost the entire temperature range~\cite{nono2019high,tchiomo2020electronic}. However, in the low temperature regions, $T\leq20$~K, and for all the samples in both device geometries, the residual resistivity is substantially the total resistivity, as the plots are all nearly straight horizontal lines. This saturation is the result of the low temperature effect in nearly pure metallic compounds as outlined by the Matthiessen's rule: To the first approximation, the residual resistivity is governed by scattering of impurities, whose contributions constitute an additive effect to the resistivity and are independent of temperature~\cite{matula1979electrical,cimberle1974deviations,bass1972deviations}. Nonetheless, deviations from the Matthiessen's rule are observed at higher temperatures ( $T>20$~K), as the temperature-dependent component of the resistivity increases in magnitude with decreasing film's thickness [see Fig.~\textcolor{blue}{S7} of the supplemental material]. Similar thickness-dependent electrical resistivity behavior was reported for MBE grown PdCoO$_2$ thin films~\cite{barbalas2022disorder}. In addition, the effective mass was found to considerably increase with decreasing film's thickness. It was suggested that, as thinner films exhibit larger surface scattering fraction and higher in-plane defects concentration, disorder-enhanced electron-phonon scattering  could be one of the possible mechanisms behind the observed deviations~\cite{barbalas2022disorder}. Note that for bulk pure and nearly pure metals such as Pd, Au, Cu, and Ag, temperature-dependent electrical resistivity analogous to that discussed here has been reported~\cite{matula1979electrical}. 

It is remarkable that the temperature minimum ($T_{\text{min}}=20$~K) at which the resistivity starts to plateau (yielding the minimum resistivity) is constant regardless of the thickness of the films and the device width [Fig.~\ref{Fig3}\textcolor{blue}{(a)} and \ref{Fig3}\textcolor{blue}{(b)}]. This suggests that the low temperatures transport properties of these films with thicknesses up to about 12~nm are mostly limited by inherent crystallographic defects, which are potentially of equal concentration~\cite{matula1979electrical,cimberle1974deviations}.  In particular, a shallow increase in the resistivity  below $T_{\text{min}}$ associated with disorder-induced localization effects in the PdCoO$_2$ films was reported in a recent study~\cite{MBrahlek_2019}. It was observed that annealing of the PdCoO$_2$ films leads to a drop in $T_{\text{min}}$ and to improved  electrical characteristics owing to the reduction of the density of defects in the films~\cite{MBrahlek_2019}.

Additionally, since the presence of impurities or defects in metals make these systems more resistive at low temperature~\cite{matula1979electrical}, the residual resistivity ratio (RRR) for a temperature close to absolute zero ($\rho_{\text{300~K}}/\rho_{\text{2~K}}$)  represents a very sensitive gauge to structural disorder in the PdCoO$_2$ films. The maximum RRR of 1.56 for our thickest PdCoO$_2$ films (11.56~nm) is two orders of magnitude smaller than that of pure bulk single crystals~\cite{THiroshi_2007}, but comparable to those recently reported for PLD and MBE grown PdCoO$_2$ films at similar thicknesses~\cite{SunJiaxin_2019,MBrahlek_2019,THarada_2018,PYordanov_2019} [Fig.~\ref{Fig3}\textcolor{blue}{(c)}]. Moreover, the RRR seems to be independent of the thickness of the films. This is contrary to the almost linear trend reported for MBE grown PdCoO$_2$ films where it was indicated that surface scattering mechanisms including scattering at the twin boundaries and film-substrate interface have a larger contribution to the low temperature electrical resistance~\cite{SunJiaxin_2019,MBrahlek_2019}.

The variation of the room temperature (RT) resistivity values as a function of the film's thickness for both device widths is plotted in Fig.~\ref{Fig3}\textcolor{blue}{(d)}. A record value of 0.85~$\mu\Omega$cm is achieved in the device with the narrower width, while a value of 2.70~$\mu\Omega$cm is obtained in the device with the wider width. The former value is one order of magnitude smaller than any value reported so far for the epitaxial growth of PdCoO$_2$ thin films~\cite{SunJiaxin_2019,MBrahlek_2019,THarada_2018}; and the latter value is somewhat equal to the RT resistivity value of bulk pure single crystals~\cite{CWHicks_2012,NNabhanila_2018,APMackenzie_2017}. It is however important to highlight that the comparatively low electrical properties of our PdCoO$_2$ thin films, especially the low RRR, stem from the presence of additional structural defects in the film such as dislocations, which originate from the large lattice mismatch between the PdCoO$_2$ film and the Al$_2$O$_3$ substrate~\cite{lee2021nonreciprocal}.

\begin{figure}[!t]
	\centering 
	\includegraphics[width=0.3\textwidth]{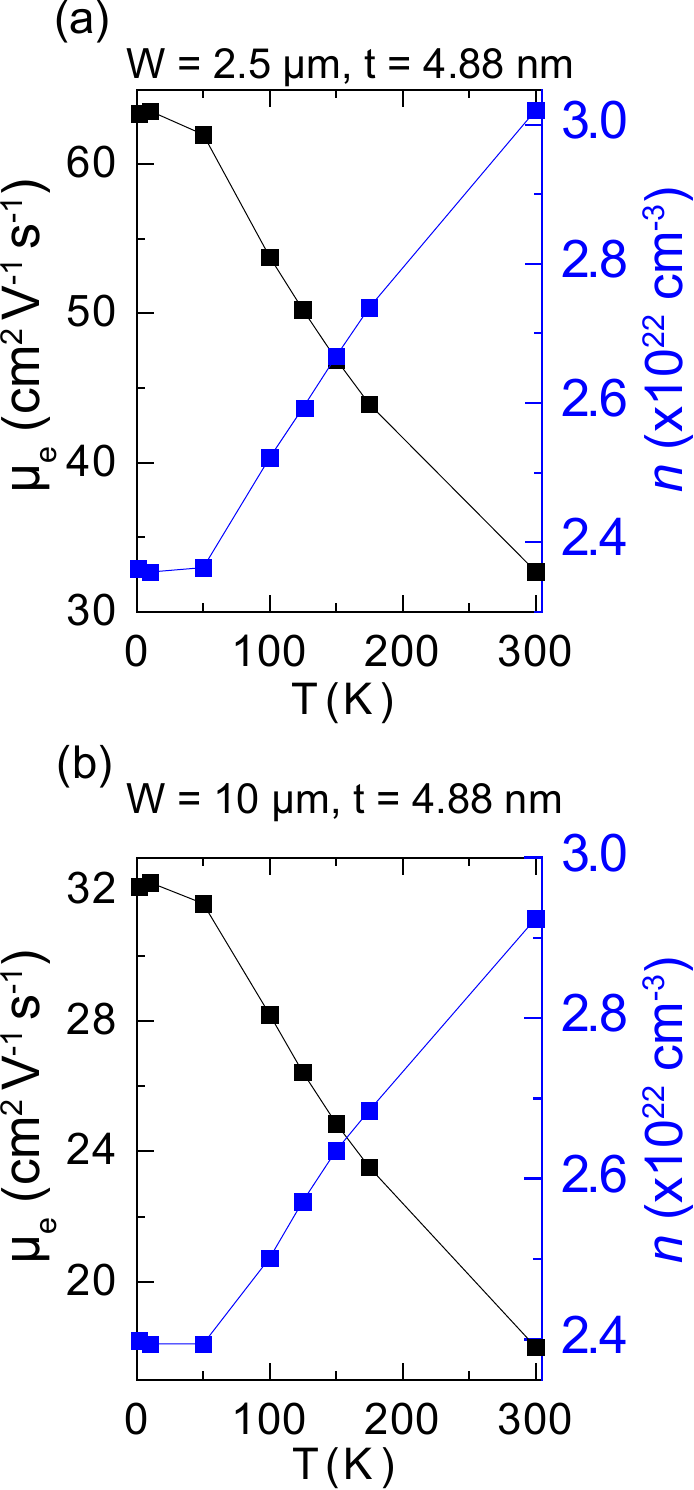}
	\caption{ \textbf{Electron mobility ($\mu_e$) and carrier density ($n$) as a function of temperature}. Electronic transport characteristics of the Hall bar devices with channel widths (a) $\text{W}=2.5~\mu$m and (b) $\text{W}=10~\mu$m fabricated on PdCoO$_2$ epitaxial films of thickness $\text{t}=4.88~\text{nm}$. These data were simultaneously extracted from Hall effect measurements with $B\perp I$, \textit{i.e.}, $\theta=0\degree$. The narrower channel $\text{W}=2.5~\mu$m yields $\mu_e$ values that are about twice those of the wider channel $\text{W}=10~\mu$m, while the \textit{n} values are comparable. \label{Fig6}}	 
\end{figure}

The magnetoresistance (MR) responses of our 4.88~nm thick PdCoO$_2$ film measured at temperatures ranging from 2~K to 300~K with the magnetic field $B\parallel c$ ($\theta=0\degree$) and oriented $90\degree$ away from $c$ ($\theta=90\degree$) are shown in Fig.~\ref{Fig4} for both Hall bar devices. It can be seen that the MR is strongly temperature dependent. When $\theta=0\degree$, ie, transverse MR [Fig.~\ref{Fig4}\textcolor{blue}{(a)} and \textcolor{blue}{(b)}], the magnetotransport presents two distinct regimes. At temperatures below 150~K , both devices exhibit negative MR which increases with decreasing temperature down to 2~K. This negative MR which is more pronounced at low temperatures ($\leq50$~K) could be thought to originate from the weak localization of the electrons in the PdCoO$_2$ sample. Weak localization is known to be an anomalous quantum phenomenon observed in the transport properties of disordered 2D systems~\cite{bergmann1984weak}. It has been reported to essentially originates from the quantum interference of the conduction electrons on the defects of the systems~\cite{bergmann1984weak}. However, when comparing the lineshapes of our negative MR data [see, Fig~\ref{Fig4}] with those of negative MR associated with the weak localization effect at comparable magnetic field strengths [see Refs.~\cite{QiSong_2022,bergmann1984weak}], it is evident that a different mechanism governs the negative MR in our PdCoO$_2$ films. 

For longitudinal MR measurement ($B\parallel I$) on a ultra-clean PdCoO$_2$ single crystal sample and for a magnetic field strength up to 9~T,  Kikugawa \textit{et al.}~\cite{kikugawa2016interplanar} observed a negative MR at 1.4~K, similar to the low temperatures MR shown in Fig~\ref{Fig4}. They argued that this negative MR could not be caused by scattering at magnetic impurities and explained in terms of weak localization effects, given the magnitude of the measured MR, the high purity as well as the non-magnetic property of the sample. This claim was supported by showing that for $B\perp I$ (transverse MR), large and positive (at all temperatures down to 1.4~K) MR could be achieved.  The authors rather attributed the observed negative MR to the emergence of the axial anomaly between the Fermi points of a field induced one dimensional electronic dispersion in PdCoO$_2$~\cite{kikugawa2016interplanar}. Given the measurement geometry of the data in Fig.~\ref{Fig4} where the current was always perpendicular to the sample \textit{c}-axis, we would believe that a different underlying quantum mechanism could be driving the negative MR in our data. Furthermore, we exclude the possibility of weak localization effects since our PdCoO$_2$ samples present no evidence of surface magnetism [see Fig.~\textcolor{blue}{S7} of the supplemental material] such as the surface ferromagnetism previously reported in an ultrathin (3.8~nm) PdCoO$_2$ film~\cite{lee2021nonreciprocal}. The absence of magnetism in our PdCoO$_2$ samples is possibly due to the mixed termination of the surface, as indicated by the relatively rough surface [see Fig.~\textcolor{blue}{S1} of the supplemental material].

It is interesting that for the same longitudinal MR measurement by Kikugawa \textit{et al.}, the negative MR is progressively suppressed by increasing the temperature, but does not completely vanishes at the highest temperature of 300~K~\cite{kikugawa2016interplanar}. This is consistent with the MR response shown in Fig.~\ref{Fig4}\textcolor{blue}{(c)} and \textcolor{blue}{(d)} for MR measurements with $B\perp c$, \textit{i.e.}, when $\theta=90\degree$. These observations suggest that the driving mechanism of the negative MR would not be suppressed when $B\perp c$, regardless of the increased thermal energy which triggers resistive mechanisms such as electron-phonon interaction. This is in contrast to the MR response presented in Fig.~\ref{Fig4}\textcolor{blue}{(a)} and \textcolor{blue}{(b)} ($\theta=0\degree$, $B\parallel c$). There, as the temperature increases, the lineshape of the MR progresses towards a more quadratic dependence from low temperatures where the MR is negative to temperatures beyond 150~K where the MR becomes positive. Positive MR is believed to be dominated by classical orbital magnetoresistive effects associated with scattering from impurities and phonons. It is remarkable that these magnetoresistive effects kick in at earlier temperatures in the thickest measured sample, $\text{t}=11.56$~nm. Positive MR are observed at temperatures $>125$~K [see Fig.~\textcolor{blue}{S3} of the supplemental material]. It is important to highlight that the temperature dependent magnetotransport properties of the 4.88~nm and 5.21~nm samples are comparable, regardless of the orientation of \textit{B} [see Fig.~\ref{Fig4}, and Fig.~\textcolor{blue}{S4} and \textcolor{blue}{S5} of the supplemental material]. Note that a similar temperature-dependent MR cross-over was recently reported in a PLD grown PdCoO$_2$ thin film~\cite{lee2021nonreciprocal}.

Figure~\ref{Fig6} depicts the evolution with temperature of the electron mobility, $\mu_e$, and the carrier density, \textit{n}, of the two Hall bar with indicated channel widths for the same PdCoO$_2$ sample studied in Fig.~\ref{Fig4}. Two distinct regimes can be identified for \textit{n}: A regime at temperatures between 300~K and 50~K where the carrier densities drop by $\approx20\%$ in cooling; and a regime at temperatures bellow 50~K where the carrier densities are independent of the temperature. These behaviors are consistent with previous reports on the temperature dependence of \textit{n} in epitaxial PdCoO$_2$ films~\cite{harada2021determination,lee2021nonreciprocal,barbalas2022disorder}. It is however noteworthy that the strong dependence of \textit{n} to high temperatures could be inferred to the level of disorder/defect in the film at high temperatures, which is also associated with sizable changes in the effective mass as discussed above. Hence, it could be that the Hall coefficient at high temperature for this class of material may not reflect the actual carrier density~\cite{barbalas2022disorder,NNabhanila_2018}. 

The extracted $\mu_e$ range between $\approx18$ and $\approx65$~cm$^2$V$^{-1}$s$^{-1}$ in both Hall bar devices from RT to 2~K. This range of electron mobility values is consistent with the small amplitudes of the MR at the field strength of 9~T and for the respective temperatures~\cite{shekhar2015extremely,lee2021nonreciprocal}. In the sample with $\text{t}=5.21~\text{nm}$ and for  both Hall bar channels [see Fig.~\textcolor{blue}{S8} of the supplemental material], the carrier densities also drop by $\approx20\%$ in a single regime from 300~K to 2~K. The electron mobilities vary from $\approx10$ to $\approx40$~cm$^2$V$^{-1}$s$^{-1}$ while cooling from RT down to 2~K, which are consistent with a recent report~\cite{lee2021nonreciprocal}.

In this study, we have reported the structural, microstructural and magnetotransport properties of Hall bar devices of various widths, fabricated on epitaxial PdCoO2 films of different thicknesses. We have analyzed two Hall bar geometries of width $\text{W}=2.5~\text{and}~10~\mu\text{m}$, all structured into the same  delafossite PdCoO$_2$ thin films of various thicknesses. The PLD prepared PdCoO$_2$ samples exhibit good structural quality with sharp interfaces that present no signs of atomic interdiffusion. In addition, both Hall bar devices exhibit transport characteristics, $\rho$, \textit{n}, and $\mu_e$, that are comparable to those reported in the literature. For the device geometries discussed in this work, we found that the narrower Hall bar, $\text{W}=2.5~\mu$m, displays slightly better electronic transport performances as compared to the wider 10~$\mu$m Hall bar. Furthermore, we have demonstrated that the MR properties of the PdCoO$_2$ samples are strongly temperature dependent, but show no variation with either the thickness of the film or the geometry of the Hall bar device. Moreover, we have noted a persistent negative MR  when $B\perp c$ at all temperatures from 2 to 300~K. For transverse MR,  $B\parallel c$, the negative MR is suppressed at temperatures above 150~K, and the MR progresses toward a more parabolic dependence, characteristic of classical orbital magnetoresistance. The lineshapes of our MR data suggest that some underlying quantum phenomenon, other than weak localization effects, would be at the origin of the emergence of this negative MR in our PdCoO$_2$ samples, especially because these samples are non magnetic. An attempt to unravel this would be to carry out a systematic study which combines theory and experiment. The results of this study provide additional basis for the understanding of the MR in PdCoO$_2$ which has proved to be highly dependent on the  orientation of the magnetic field, a property that could be relevant to certain applications. 

\begin{acknowledgements}
	P. Ngabonziza. acknowledges startup funding from the College of Science and the Department of Physics \& Astronomy at Louisiana State University. The authors acknowledge Dr Julia Deuschle for the FIB sample preparation for the STEM investigations.
	
\end{acknowledgements}

\bibliography{references_PdCoO2_Paper}
\onecolumngrid
\newpage
\setcounter{table}{0}
\setcounter{figure}{0}
\renewcommand{\thefigure}{S\arabic{figure}}%
\setcounter{equation}{0}
\renewcommand{\theequation}{S\arabic{equation}}%
\setstretch{1.25}

\begin{center}
	\title*{\textbf{\large{Supplementary material}:\\ [0.25in]	 \large{Magnetotransport properties in epitaxial films of metallic delafossite PdCoO$_2$: Effects of thickness and width variations in Hall bar devices}}}
\end{center}

\begin{center}
	\small{Arnaud P. Nono Tchiomo$^{1)}$, Anand Sharma$^{1)}$, Sethulakshmi Sajeev$^{1)}$, Anna Scheid$^{2)}$,} \small{Peter A. van Aken$^{2)}$, Takayuki Harada$^{3)}$, Prosper Ngabonziza$^{1,4)}$}\\ 
	\small{$^{1)}$\textit{Department of Physics and Astronomy, Louisiana State University, Baton Rouge, LA 70803, USA}}\\ \small{$^{2)}$\textit{Max Planck Institute for Solid State Research, Heisenbergstr. 1, 70569 Stuttgart, Germany}}\\
	\small{$^{3)}$\textit{Research Center for Materials Nanoarchitectonics (MANA), National Institute for}}\\ 
	\small{\textit{ Materials Science, Tsukubashi, Ibaraki 305-0044, Japan}}\\
	\small{$^{4)}$\textit{Department of Physics, University of Johannesburg, P.O. Box 524 Auckland Park 2006, Johannesburg, South Africa}}
\end{center}

\section{Design of the Hall bar devices}

PdCoO$_2$ thin films of various thicknesses were epitaxially grown on Al$_2$O$_3$ substrates using optimized deposition conditions to ensure high crystallinity and uniformity across each sample. These films were then patterned into Hall bar devices with channel widths ranging from 500~nm to 10~$\mu$m, allowing us to investigate the impact of device geometry on electronic and magnetotransport properties. Standard optical lithography (MW3) was employed to achieve precise patterning, resulting in well-defined Hall bar structures that enabled reliable transport measurements. To establish stable electrical contacts, each device was designed with Ti and Au layers of 5 and 45~nm, respectively, deposited via electron-beam evaporation. This provided robust charge injection capabilities suitable for measurements across a broad temperature range. Top-view scanning electron microscopy (SEM) verified the patterning accuracy and the structural integrity of the devices, while a side-view schematic of the devices illustrated the arrangement of the PdCoO$_2$ films on the Al$_2$O$_3$ substrate (see main text). Aluminum wire bonding was used to establish durable connections at the device corners, which remained reliable throughout extensive temperature cycling and repeated measurements.

\section{STEM data acquisition}

Electron-transparent TEM specimens of the sample were prepared on a FEI Scios focused ion beam (FIB) using the standard lift-out method. Samples with a size of $20\times5~\mu m^2$ were thinned to 30~nm with 5~kV Ga ions, followed by a final polish at 2~kV to reduce the effects of surface damage.

STEM investigations were performed using a JEOL JEM-ARM200F equipped with a cold field-emission gun and a probe Cs corrector (DCOR, CEOS GmbH). The measurements were performed at ambient temperature with an acceleration voltage of 200~kV. In order to improve the signal-to-noise ratio, reduce scanning artifacts, and address sample drift effects, the STEM images were generated from multi-frame acquisitions using high scanning speeds and applying post-acquisition cross-correlation. In addition, strategic rotation of the scan window with respect to the atomic lattice was implemented to avoid superposition of lattice peaks in the Fourier transform with features associated with scan artifacts. 

The EELS data were acquired using a Gatan GIF Quantum ERS imaging filter with a 5~mm entrance aperture and a 1.5~cm camera length, resulting in a collection semi-angle of 111~mrad. Principal component analysis (PCA) was used to improve the signal-to-noise ratio, and 15 principal components were used for accurate elemental mapping~[1]. 

\newpage

\section{AFM and XRD rocking curves}


\begin{figure*}[!h]
\centering
	{\includegraphics[width=0.75\textwidth]{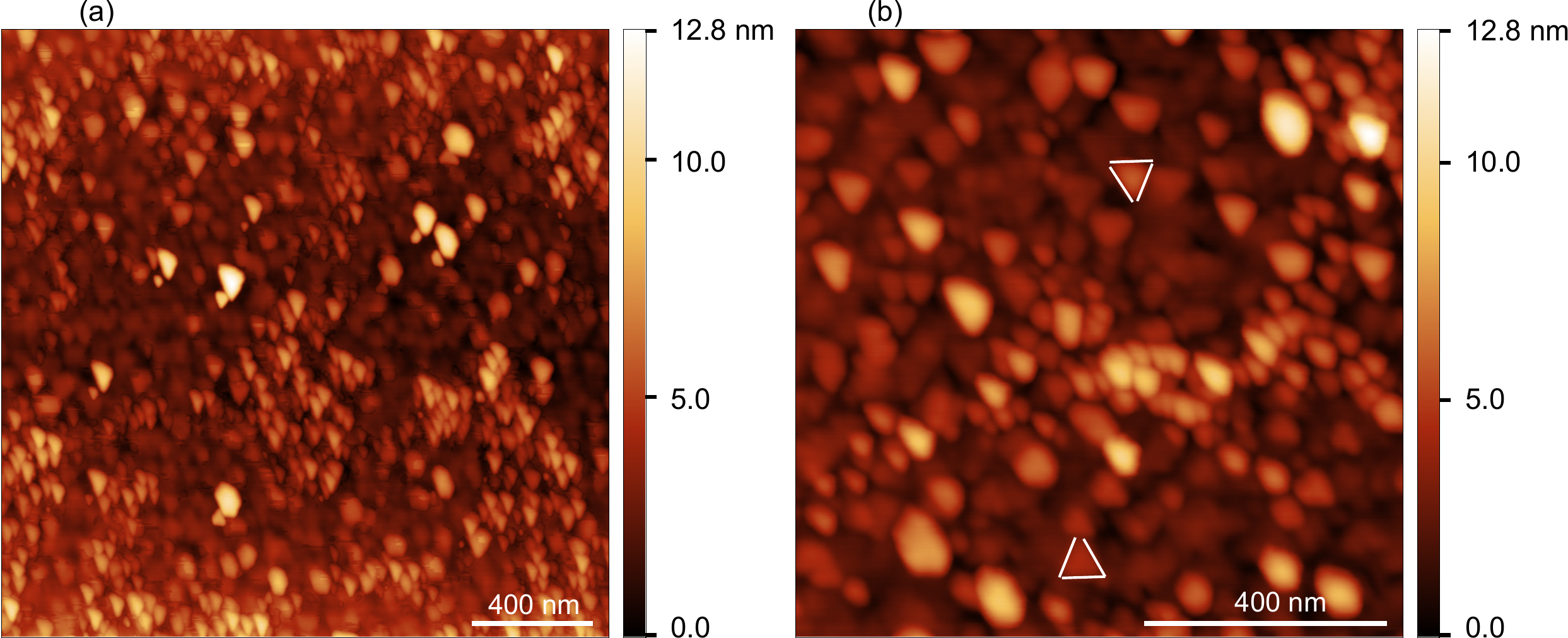}}  \\       
	\textbf{Figure S1:} Atomic force microscopy (AFM) images of a representative PdCoO$_2$ sample. (a) Triangular structures are visible at the surface. (b)  Zoomed AFM image depicting two triangular features rotated 180$\degree$ from one another. This is an evidence of the existence of twin domains in the PdCoO$_2$ film.
\end{figure*}

\begin{figure*}[!h]
	{\includegraphics[width=0.75\textwidth]{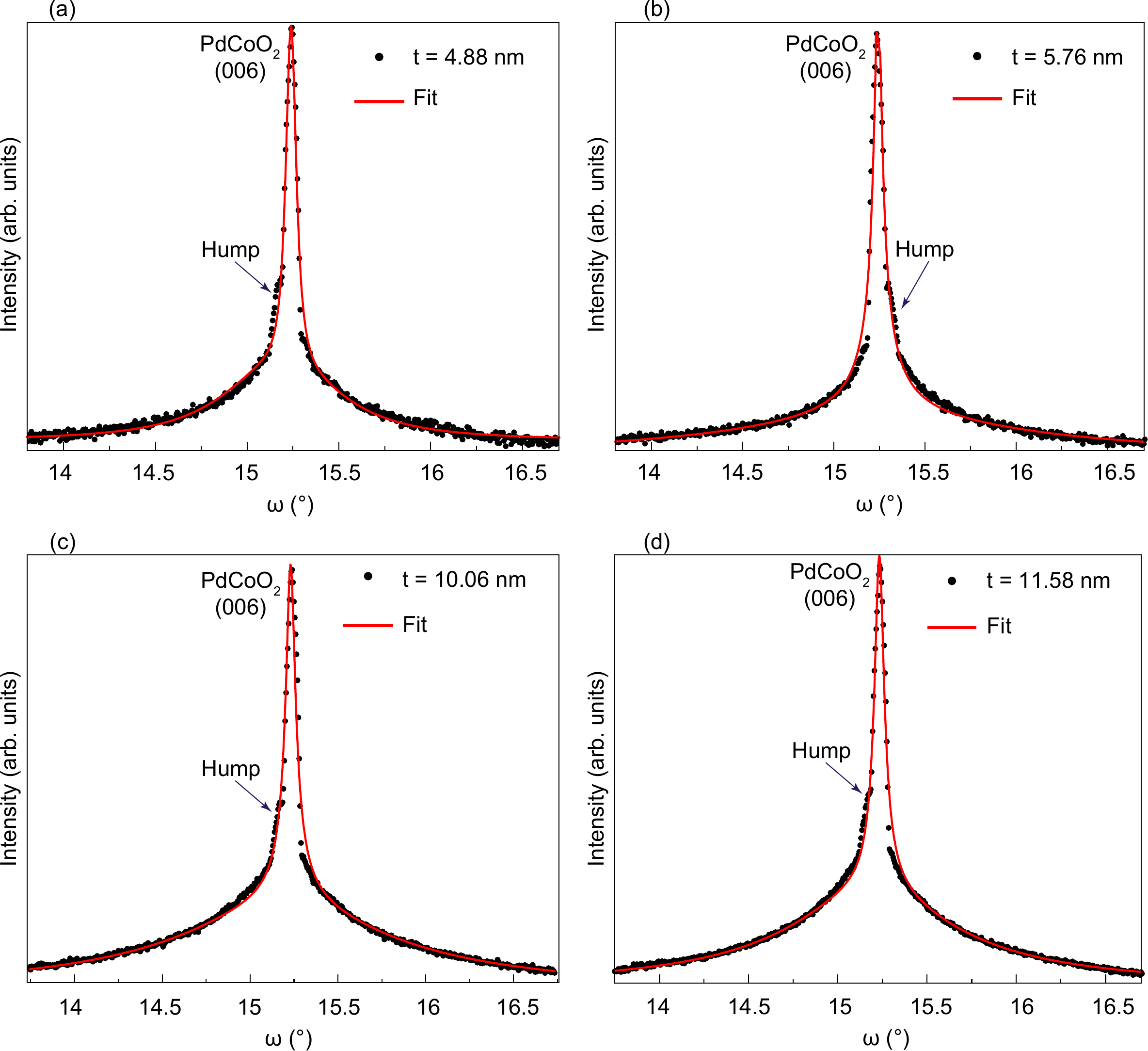}}  \\       
	\textbf{Figure S2:} Fitted rocking curves around the (006) reflections of various PdCoO$_2$ films with different thicknesses (t). The full width at falf maximum (FWHM) values plotted in Fig~\textcolor{blue}{1(f)} of the main text were extracted from these fits. Note the consistent presence of a small hump in about 0.5$\degree$ interval around the film peak.
\end{figure*}

\clearpage

\section{Temperature and angle dependent magnetotransport}

\begin{figure*}[!h]
	\begin{center}
		{\includegraphics[width=0.65\textwidth]{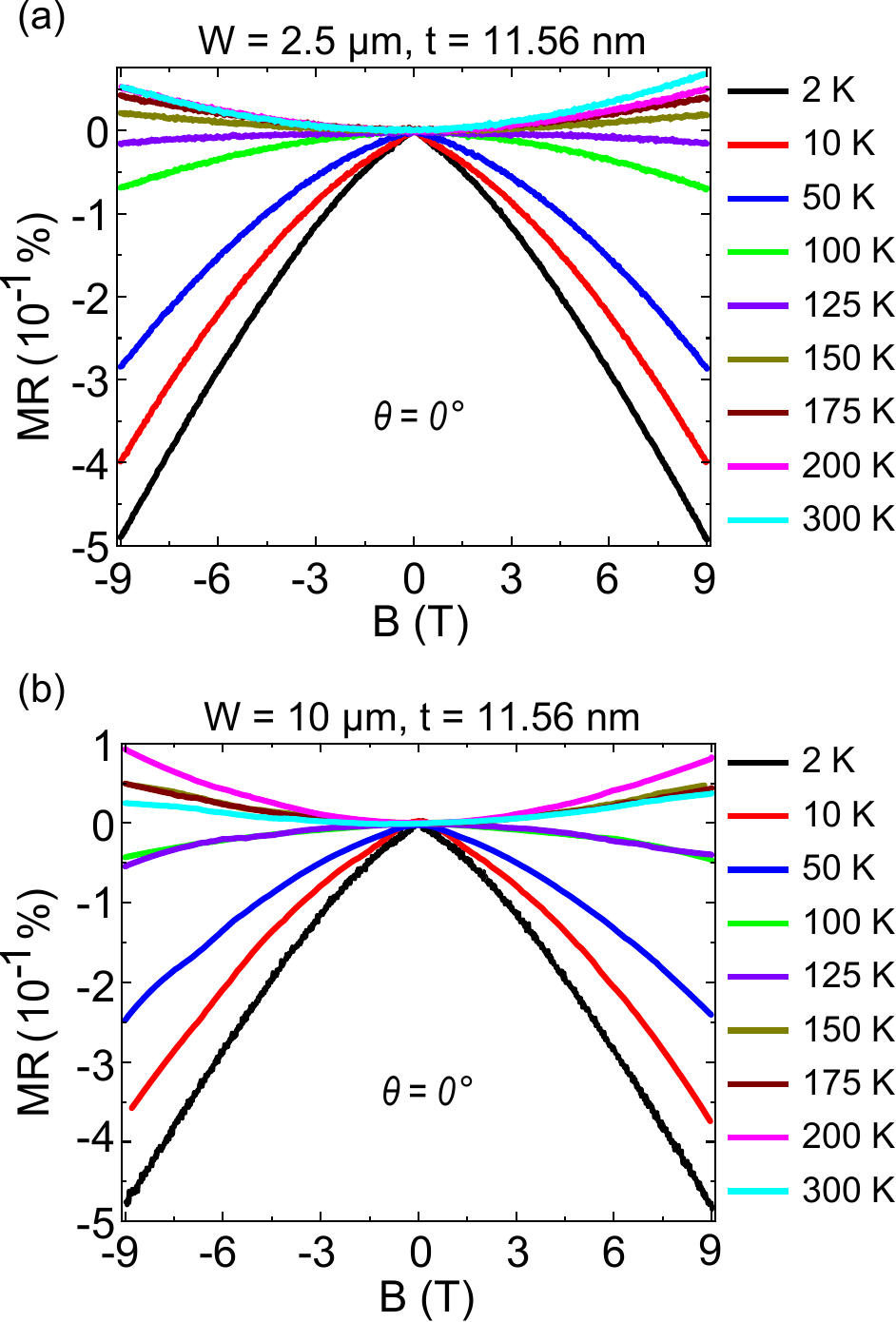}}   
	\end{center}
	\textbf{Figure S3:} Temperature-dependent magnetoresistance data for the  11.56~nm thick sample, measured simultaneously from the (a) 2.5~$\mu$m and (b) 10~$\mu$m Hall bar devices with the magnetic field \textit{B} parallel to the sample \textit{c}-axis. Note here that the MR becomes positive at the temperature of 125~K, compared to 150~K in the thinner 4.88 and 5.26~nm samples.
\end{figure*}

\begin{figure*}[!h]
	\begin{center}
		{\includegraphics[width=0.65\textwidth]{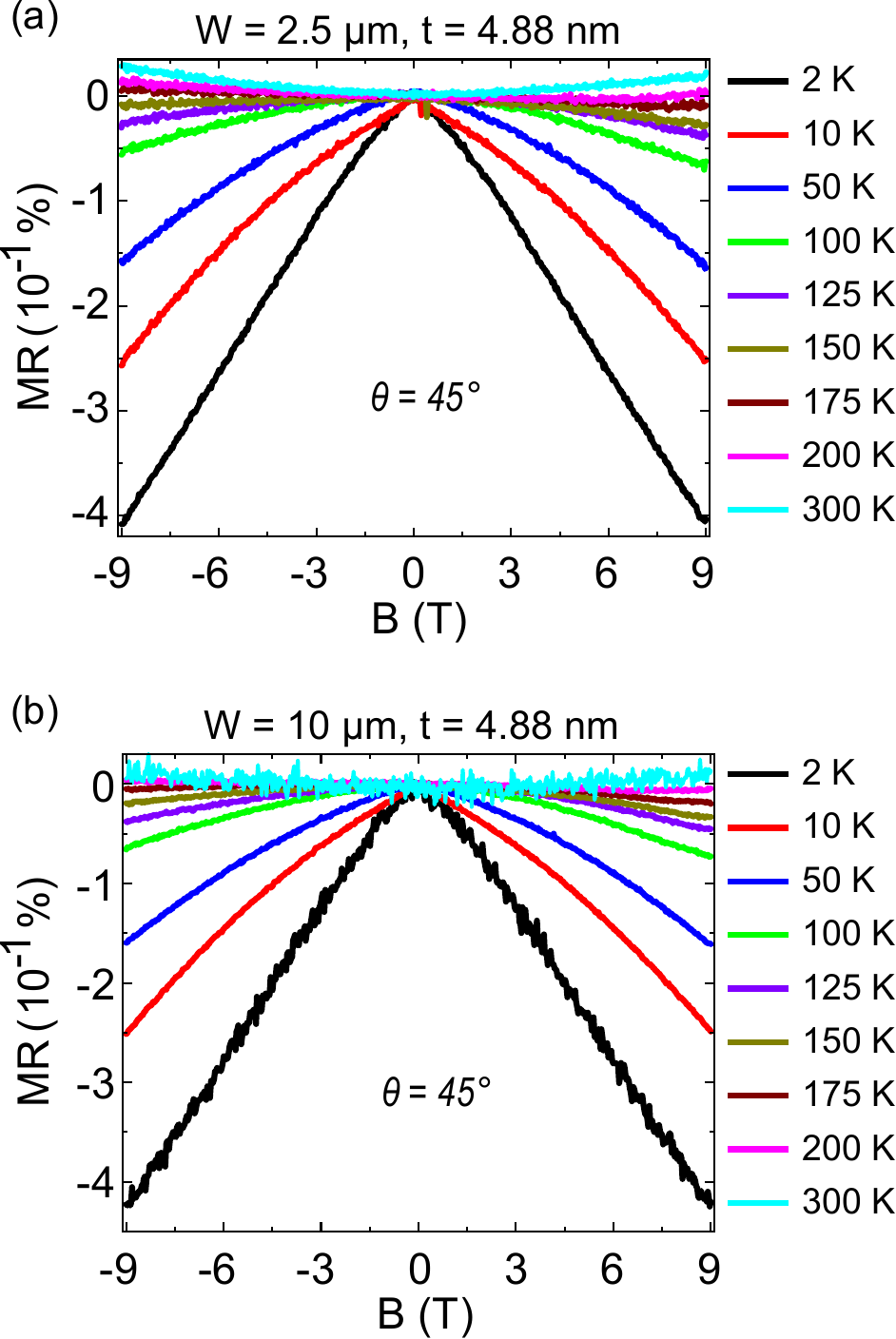}}   
	\end{center}
	\textbf{Figure S4:} Temperature-dependent magnetoresistance data for the 4.88~nm sample, measured with the magnetic field \textit{B} oriented 45$\degree$ away from the sample \textit{c}-axis. The data are acquired simultaneously from the (a) 2.5~$\mu$m and (b) 10~$\mu$m Hall bar devices. The MR data for this sample are the same whether \textit{B} is 45$\degree$ away from \textit{c} or $\parallel$ to \textit{c} (see main text). 
\end{figure*}

\begin{figure*}[!t]
	\begin{center}
		{\includegraphics[width=1\textwidth]{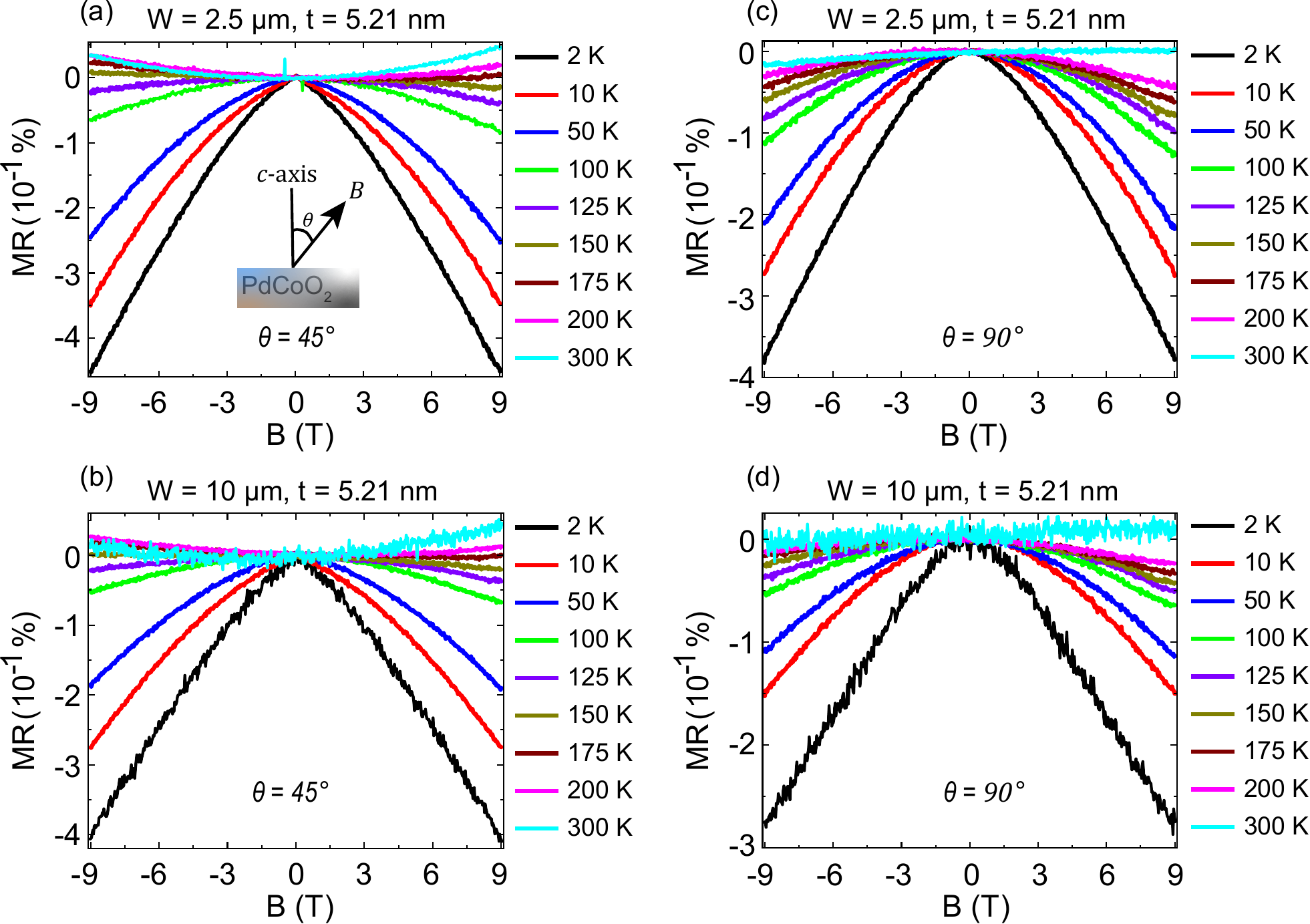}}   
	\end{center}
	\textbf{Figure S5:} Temperature-dependent magnetoresistance data for the 5.21~nm sample measured with the magnetic field \textit{B} oriented (a) and (b) 45$\degree$, and (c) and (d) 90$\degree$ away from the sample \textit{c}-axis. The data were measured from the  2.5~$\mu$m and 10~$\mu$m Hall bar devices, simultaneously in (a) and (b), and (c) and (d), respectively. Note the similarity between the data for both 4.88~nm and 5.21~nm samples, on both Hall bar devices and at the respective orientation of \textit{B}.
\end{figure*}

\clearpage

\section{Electronic transport}

\begin{figure}[!h]
	\begin{center}
		{\includegraphics[width=0.75\textwidth]{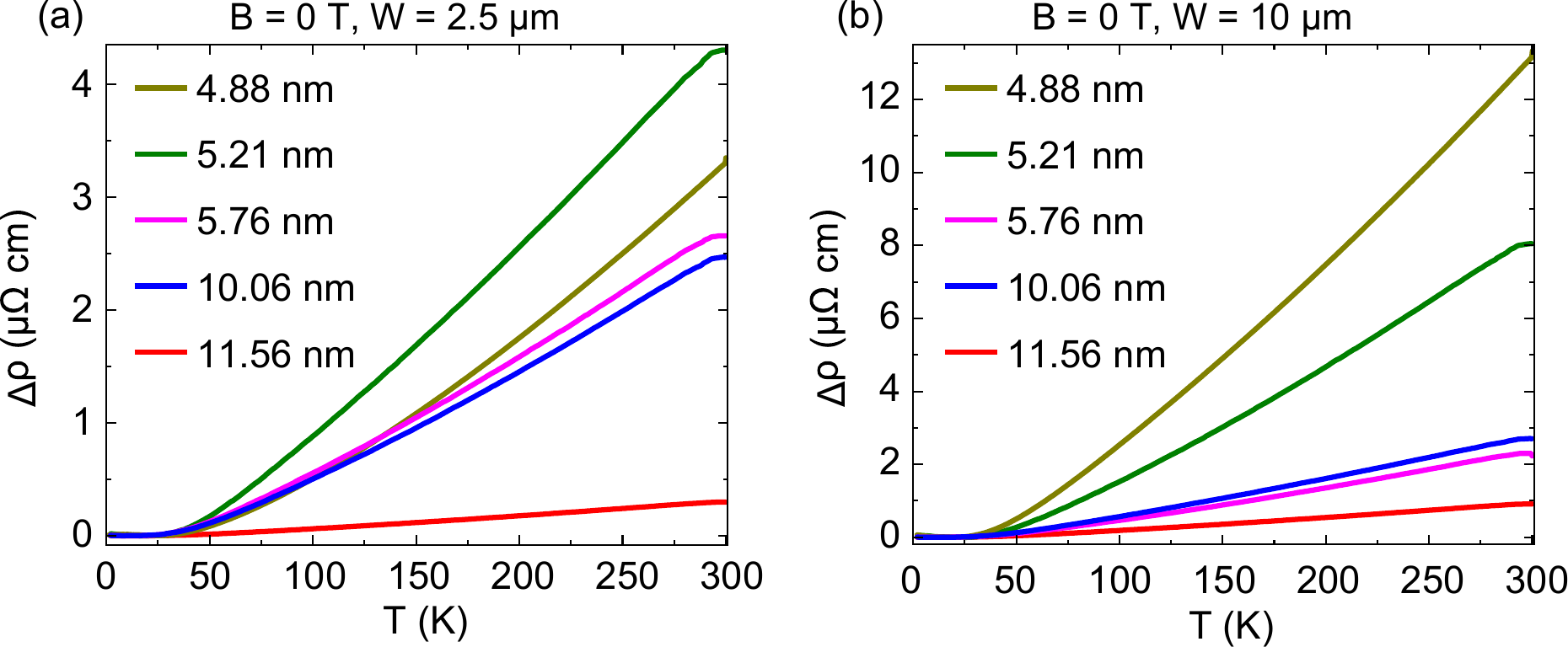}}
	\end{center}    
	\textbf{Figure S6:} Relative temperature-dependence $\Delta\rho(T)=\rho(T)-\rho_0$ (with $\rho_0=\rho(2~\text{K})$) of the resistivity   for two Hall bar devices structured onto PdCoO$_2$ thin films of various thicknesses. This shows how the temperature-dependent part of the resistivity increases in magnitude as the thickness of the film decreases.
\end{figure}

\begin{figure}[!h]
	\begin{center}
		{\includegraphics[width=0.65\textwidth]{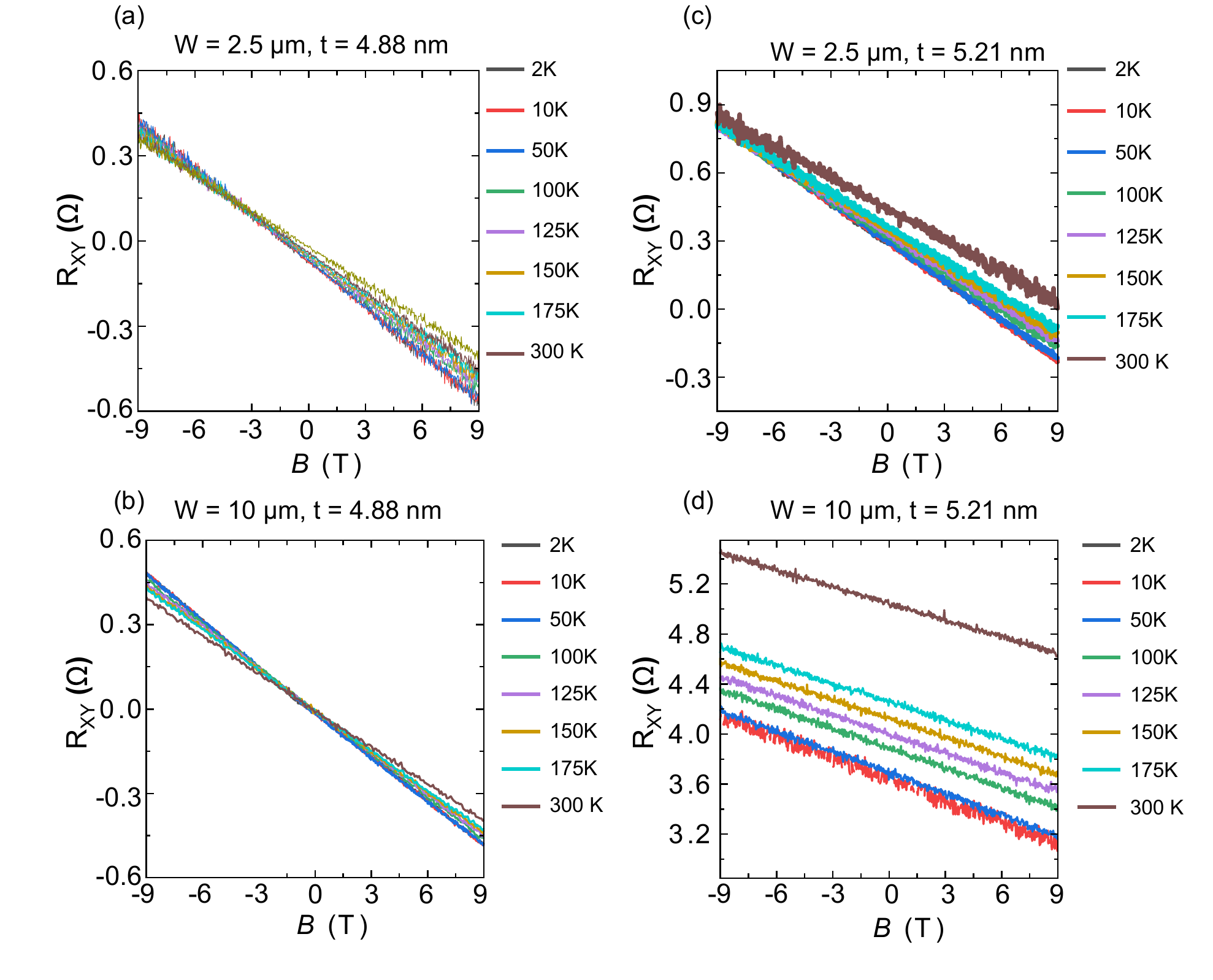}}
	\end{center}    
	\textbf{Figure S7:} Temperature-dependent Hall resistivity $\rho_{xy}(B)$  measurements for two Hall bar devices structured onto PdCoO$_2$ thin films of thicknesses (a) and (b) 4.88~nm, and (c) and (d) 5.21~nm. The data were simultaneously acquired from the channel $\text{W}=2.5$ and $10~\mu\text{m}$ in (a) and (b), and (c) and (d), respectively. All the Hall measurements were acquired with the magnetic field $B$ applied parallel to the crystallographic \textit{c} axis, i.e., perpendicular to the PdCoO$_2$  conducting planes. Note the linear slope of $\rho_{xy}(B)$ at all temperatures when \textit{B} is swept from -9 to 9~T, indicating no sign of magnetism at the surface of the films~[2].
\end{figure}

\begin{figure*}[!h]
	\begin{center}
		{\includegraphics[width=0.8\textwidth]{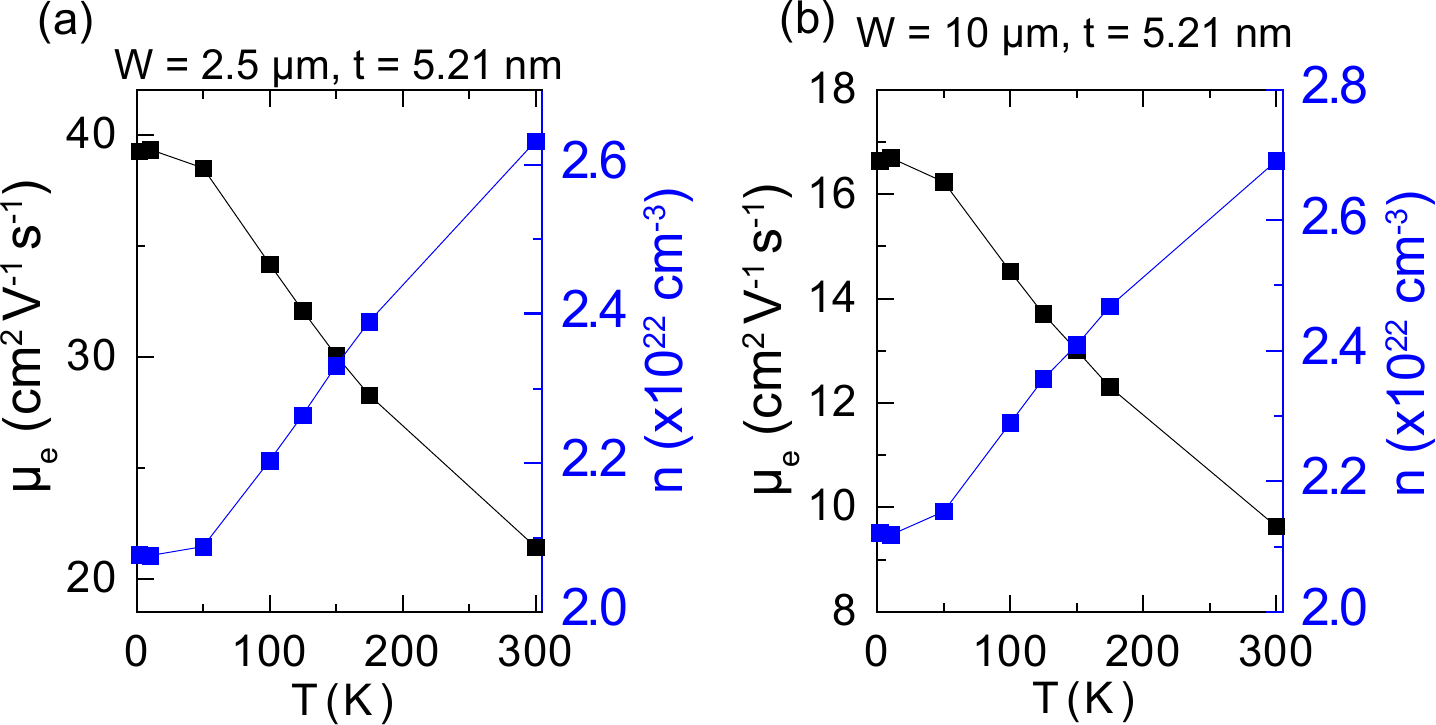}}   
	\end{center}
	\textbf{Figure S8:} Temperature-dependent carrier densities and electron mobilities for the 5.21~nm thick PdCoO$_2$ sample. These data were simultaneously extracted from the Hall data shown in Fig.~\textcolor{blue}{S4(c)} and \textcolor{blue}{(d)} above. 
\end{figure*}


\newpage
\section{References}

	[1]. Cueva, P., Hovden, R., Mundy, J. A., Xin, H. L. \& Muller, D. A. \textit{Data processing for atomic resolution 
	electron energy loss spectroscopy. Microscopy and microanalysis} : the official journal of Microscopy Society of America, Microbeam Analysis Society, Microscopical Society of Canada \textbf{18}, 667~-~675; 10.1017/S1431927612000244 (2012).

	[2].  J. H. Lee, T. Harada, F. Trier, L. Marcano, F. Godel, S. Valencia, A. Tsukazaki, and M. Bibes, Nano Lett. \textbf{21}, 8687 (2021).

\end{document}